\documentclass[reprint,superscriptaddress,amsmath,amssymb,aps,prd,notitlepage,longbibliography,floatfix,nofootinbib]{revtex4-1}

\usepackage{graphicx}   
\usepackage[
colorlinks=true,        
citecolor=blue,         
linkcolor=blue,         
urlcolor=blue           
]{hyperref}             
\usepackage{bm}         
\usepackage{xcolor}     
\usepackage{array}      
\usepackage{lipsum}
\usepackage{color}      
\usepackage[section]{placeins} 
\usepackage{makecell}

\newcommand{\Eq}[1]{Eq.~\eqref{#1}}     
\newcommand{\Fig}[1]{Fig.~\ref{#1}}     
\newcommand{\Table}[1]{Table~\ref{#1}}  

\newcommand{\dif}{\mathrm{d}}
\newcommand{\eBasis}{\bm{e}}
\newcommand{\eConst}{\mathrm{e}}
\newcommand{\iImag}{\mathrm{i}}
\newcommand{\vecOmega}{\bm{\Omega}}

\begin{document}
	
\title{Upper limits on the  Polarized Isotropic Stochastic  Gravitational-Wave Background from Advanced LIGO-Virgo's First Three Observing Runs}
	
\author{Yang Jiang}
\affiliation{CAS Key Laboratory of Theoretical Physics, 
		Institute of Theoretical Physics, Chinese Academy of Sciences,
		Beijing 100190, China}
\affiliation{School of Physical Sciences, 
		University of Chinese Academy of Sciences, 
		No. 19A Yuquan Road, Beijing 100049, China}
\author{Qing-Guo Huang}
\email{Corresponding author: huangqg@itp.ac.cn}
\affiliation{CAS Key Laboratory of Theoretical Physics, 
		Institute of Theoretical Physics, Chinese Academy of Sciences,
		Beijing 100190, China}
\affiliation{School of Physical Sciences, 
		University of Chinese Academy of Sciences, 
		No. 19A Yuquan Road, Beijing 100049, China}
\affiliation{School of Fundamental Physics and Mathematical Sciences
		Hangzhou Institute for Advanced Study, UCAS, Hangzhou 310024, China}

\date{\today}
	
\begin{abstract}

Parity violation is expected to generate an asymmetry between the amplitude of left and right-handed gravitational-wave modes which leads to a circularly polarized stochastic gravitational-wave background (SGWB). Due to the three independent baselines in the LIGO-Virgo network, we focus on the amplitude difference in strain power characterized by Stokes' parameters and do maximum-likelihood estimation to constrain the polarization degree of SGWB. Our results indicate that there is no evidence for the circularly polarized SGWB in the data. Furthermore, by modeling the SGWB as a power-law spectrum, we place upper limit on the normalized energy density  $\Omega_\text{gw}(25\,\text{Hz})<5.3\times10^{-9}$ at $95\%$ confidence level after marginalizing over the polarization degree and spectral index.
\end{abstract}
	
\maketitle
{\it Introduction. }
The stochastic gravitational-wave background (SGWB) is a superposition of gravitational waves (GWs) from numerous unresolved and uncorrelated sources. These GW sources can arise either from  astrophysical processes like compact binary coalescences (CBCs) \cite{PhysRevD.84.084004,PhysRevD.85.104024,Zhu:2012xw}, rotating neutron stars \cite{Ferrari:1998jf,Regimbau:2001kx,PhysRevD.87.063004}, stellar core collapses \cite{Crocker:2015taa}, or cosmological contributions, such as phase transitions (PT) \cite{PhysRevD.30.272,Kosowsky:1992rz,Dev:2016feu,VonHarling:2019rgb}, cosmic strings \cite{PhysRevLett.85.3761,Sarangi:2002yt,PhysRevLett.98.111101,LIGOScientific:2017ikf} and inflation models \cite{PhysRevD.55.R435,Guzzetti:2016mkm}. The detection of SGWB will provide a better understanding about the distribution of astrophysical sources, the history of early Universe and testing the theories of gravity. Until now, the terrestrial laser interferometers like Advanced LIGO \cite{AdvancedLIGO2015} and Virgo \cite{VIRGO:2014yos} have been combined to search for SGWB in their output strain data. Based on current noise power, the results show that there is no detectable correlation and thus upper limits on the energy density of both isotropic and anisotropic SGWB have been set in  \cite{KAGRA:2021kbb,KAGRA:2021mth,LIGOScientific:2021qeg}.

The SGWB is widely assumed to be unpolarized, implying that the different polarized GW modes are treated as statistically identical and independent in the analyses. This assumption is quite reliable for a superposition of signals from clustered sources and parity is conserved in general relativity (GR). However, once the parity is violated, the asymmetry between the amplitude of left and right-handed GW modes leads to a circularly polarized  SGWB. Some processes in the early Universe, such as helical turbulence during a first-order PT \cite{Kahniashvili:2005qi}, can generate a polarized SGWB. When we go beyond GR, various theories of gravity, including Chern-Simons gravity \cite{Alexander:2009tp,Lyth:2005jf,Satoh:2007gn,Saito:2007kt}, Ho\v{r}ava-Lifshitz gravity \cite{Horava:2009uw,PhysRevLett.102.231301} and ghost-free scalar-tensor gravity \cite{Crisostomi:2017ugk} etc, can also give rise to parity violations. Accounting for these theories, particular polarization mode can be enhanced or compressed through amplitude birefringence effect \cite{Zhao:2019xmm} during its propagation, resulting in unequal left and right-handed components in SGWB. In all, detecting a circularly polarized SGWB might yield a profound discovery for fundamental physics. 

In this letter, we perform the first multi-baseline search for the circularly polarized isotropic SGWB in the data of Advanced LIGO--Virgo's First Three Observing Runs. The method for detecting the circularly polarized SGWB by ground-based interferometers was proposed for the first time in \cite{Seto:2007tn,Seto:2008sr}. This method depends on the detector's nontrivial reaction to the Stokes V parameter. Even though SGWB can be detected in a one-baseline analysis, the existence of polarization cannot be affirmed because one baseline cannot distinguish the left-handed component from the right-handed one in principle. In this sense, in \cite{Crowder:2012ik}, the authors can obtain the upper limit on parity violation indirectly by assuming a fiducial model with a power-law spectrum of SGWB and then detecting the  deviation, because only the correlation of LIGO Hanford-LIGO Livingston pair was available at that time. In principle, there are two different polarization modes, at least one more baseline is needed for detecting the circularly polarized SGWB. Fortunately, from O3 observing run, Virgo has been involved in searching for SGWB. In this letter, the extra outputs of LIGO-Virgo pairs are firstly adopted for distinguishing  different polarization modes in the data.


\bigskip

{\it Circularly Polarized SGWB. }
The SGWB is formed as a superposition of plane waves propagating along all possible directions with various frequencies:
\begin{equation}
    h_{ij}(t,\bm{x})=\sum_{A}\int_{-\infty}^{\infty}\dif f\int\dif^2\Omega\; h_A(f,\bm{\Omega})\eConst^{2\pi\iImag f(t-\bm{\Omega}\cdot\bm{x}/c)}\eBasis_{ij}^A(\bm{\Omega}),
\end{equation}
where $\eBasis_{ij}^A$ is the polarization basis tensor. As far as we are concerned, the circularly polarization basis $A=\{R,L\}$ is a favorable choice which is related to the linearly polarization basis by 
\begin{equation}
    \eBasis_{ij}^R=\frac1{\sqrt2}\left(\eBasis^+_{ij} + \iImag\eBasis^\times_{ij}\right),\;\eBasis_{ij}^L=\frac1{\sqrt2}\left(\eBasis^+_{ij} - \iImag\eBasis^\times_{ij}\right).
\end{equation}
For isotropic SGWB, the quadratic expectation values of $h_A$ is
\begin{equation}
    \langle h_A(f,\bm{\Omega})h^*_{A^\prime}(f^\prime,\bm{\Omega}^\prime) \rangle=\frac12 S_h^{A}(f)\delta_{AA^\prime}\delta(f-f^\prime)\frac{\delta^2(\vecOmega,\vecOmega^\prime)}{4\pi},
\end{equation}
where
\begin{equation}
    S_h^R=\frac12\left(I(f)+V(f)\right),\quad S_h^L=\frac12\left(I(f)-V(f)\right),  \label{eq:stokes}
\end{equation}
and $I(f),V(f)$ are the so-called Stokes' parameters. To reveal their meanings, for a sinusoidal plane wave, the concise expressions for both of them are given by  
\begin{equation}
    I=|h_R|^2+|h_L|^2,\quad V=|h_R|^2-|h_L|^2.
\end{equation}
In this sense, $I(f)$ and $V(f)$ denote the total intensity of SGWB and the degree of parity violation, respectively. If $V\neq 0$, the parity is violated. 
The normalized energy density of SGWB is defined by 
\begin{equation}
    \Omega_\text{gw}(f)=\frac{1}\rho_c\frac{\dif \rho_\text{gw}}{\dif \ln f},
\end{equation}
where $\rho_c=3H_0^2c^2/(8\pi G)$ is the critical energy density and $\Omega_\text{gw}$ is related to the Stokes' parameter $I(f)$ by 
\begin{equation}
    \Omega_\text{gw}(f)=\frac{2\pi^2 f^3}{3H_0^2}I(f). \label{eq:IOmega}
\end{equation}

\bigskip

\begin{figure}
    \centering
    \includegraphics[width=\columnwidth]{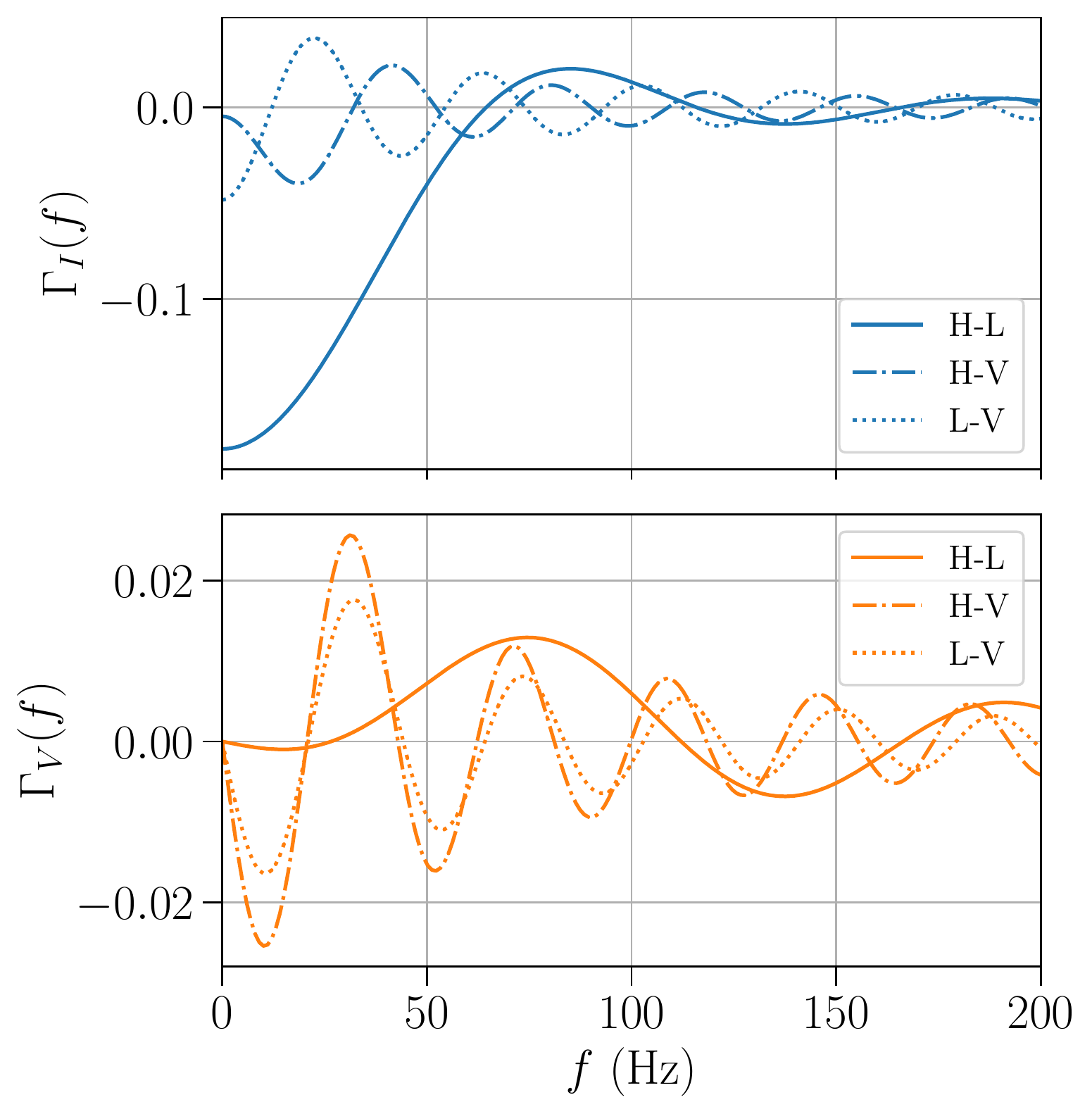}
    \caption{Overlap reduction functions of H-L (LIGO Hanford-LIGO Livingston), H-V (LIGO Hanford-Virgo) and L-V (LIGO Livingston-Virgo) pairs, respectively. }
    \label{fig:overlap}
\end{figure}

{\it Separating different circularly polarized modes. }
The detection of SGWB relies on the correlation analysis \cite{Allen:1997ad} of strain data. If two detectors couple together to form a baseline, during a certain period $T$, a cross-correlating statistic is defined as
\begin{equation}
    \hat{C}(f)=\frac2T \mathrm{Re}\left[\tilde{s}_1^*(f)\tilde{s}_2(f)\right], 
\end{equation}
where $\tilde{s}_{1,2}(f)$ are the Fourier transform of the time-series strain outputs, $\hat{C}(f)$ is expected to be Gaussian distributed with expectation
\begin{equation}
    \langle \hat{C}(f) \rangle=\Gamma_I(f)I(f)+\Gamma_V(f)V(f), \label{eq:Cmean}
\end{equation}
and variance (small signal-to-noise ratio limit)
\begin{equation}
    \sigma^2(f)\approx \frac{1}{2T\Delta f}P_1(f)P_2(f).  \label{eq:sigma}
\end{equation}
Here $\Delta f$ is the width of frequency bins and $P_{1,2}(f)$ are the one-side noise power spectrum of the detectors. $\Gamma(f)$ is the overlap reduction function of the baseline. The overlap reduction function can be calculated by the antenna pattern of the detector \cite{Crowder:2012ik} as follows 
\begin{equation}
    \begin{split}
         \Gamma_I(f) &=\frac{1}{8\pi}\int\dif^2\Omega \left(F_1^+F_2^{+*}+F_1^\times F_2^{\times*}\right)\eConst^{2\pi\iImag f\vecOmega\cdot\Delta\bm{x}}, \\
         \Gamma_V(f) &=-\frac{\iImag}{8\pi}\int\dif^2\Omega \left(F_1^+F_2^{\times*}-F_1^\times F_2^{+*}\right)\eConst^{2\pi\iImag f\vecOmega\cdot\Delta\bm{x}}. 
    \end{split}
\end{equation}
Note that there are three baselines denoted by H-L (LIGO Handford-LIGO Livingston), H-V (LIGO Hanford-Virgo) and L-V (LIGO Livingston-Virgo), respectively. The  overlap functions of the interferometer pairs involved in our analysis are plotted in \Fig{fig:overlap}. 

In this letter we adopt the data released by LVK Collaborations during O1$\sim$O3 observing runs \cite{LVK:IsoSGWBdata}. The analyzed frequency band is $20\sim1726$ Hz with a  resolution of $1/32$ Hz. For the case with more than one baseline, the combined likelihood can be written as 
\begin{equation}
    p(\bm{\hat{C}}|S)\propto \prod_{f}\exp\left[-\frac{1}{2}\left(\bm{\hat{C}}-\Gamma S\right)^\dagger N^{-1}\left(\bm{\hat{C}}-\Gamma S\right)\right].
\label{likelihoodc}
\end{equation}
The noise correlation matrix $N$ is diagonal with elements in \Eq{eq:sigma} and
\begin{equation}
    \bm{\hat{C}}=\begin{bmatrix}\hat{C}_1\\\hat{C}_2\\ \vdots\end{bmatrix},\:
    \Gamma=\begin{bmatrix} \Gamma_{1I} & \Gamma_{1V}\\\Gamma_{2I} & \Gamma_{2V}\\\vdots & \vdots \end{bmatrix},\:
    S=\begin{bmatrix} I\\V \end{bmatrix}.
\end{equation}
Maximizing the likelihood is equivalent to minimizing
\begin{equation}\Vert \bm{\hat{C}}^\prime-\Gamma^\prime S \Vert, \end{equation} 
where $\bm{\hat{C}}^\prime=\sqrt{N^{-1}}\bm{\hat{C}}$ and $\Gamma^\prime=\sqrt{N^{-1}}\Gamma$. This least--squares problem can be solved by applying singular value decomposition \cite{2009PhRvD..80l2002T,Press2007NumericalRT} to $\Gamma^\prime$
\begin{equation}
    \Gamma^\prime=U\Sigma W^\dagger, 
\end{equation}
here $U$ and $W$ are unitary matrices, $\Sigma$ is diagonal with singular values $w_i$ arranged from large to small. This decomposition defines the \emph{Moore-Penrose inverse}
\begin{equation}
    \Gamma^{\prime-1}=W\cdot\mathrm{diag}(1/w_i)\cdot U^\dagger. \label{eq:pseuinv}
\end{equation}
The maximum likelihood estimator is
\begin{equation}
    \hat{S}=\Gamma^{\prime-1}\hat{\bm{C}}^\prime,
\end{equation}
and the covariance becomes
\begin{equation}
    \mathrm{Cov}_{ij}=\sum_k\frac{W_{ik}W_{jk}}{w_k^2}.
\end{equation}
Here we need to remind that the magnitude of $w_i$ is a criterion of matrix singularity. When $w_i\to0$, the covariance matrix tends to be infinite. This happens quite often due to the degeneration of overlap functions and glitches in detector's frequency band. 
Actually, these insensitive frequency bins not only make little contribution to the constraint on energy spectrum, but also make trouble to numerical calculation. Therefore, in practice, the frequency bins with $w_i/w_\text{max}<10^{-10}$ will be removed in our analysis. This criteria cuts off about $18.9\%$ frequency bins of $\bm{\hat{C}}(f)$. 

\begin{figure}
    \centering
    \includegraphics[width=\columnwidth]{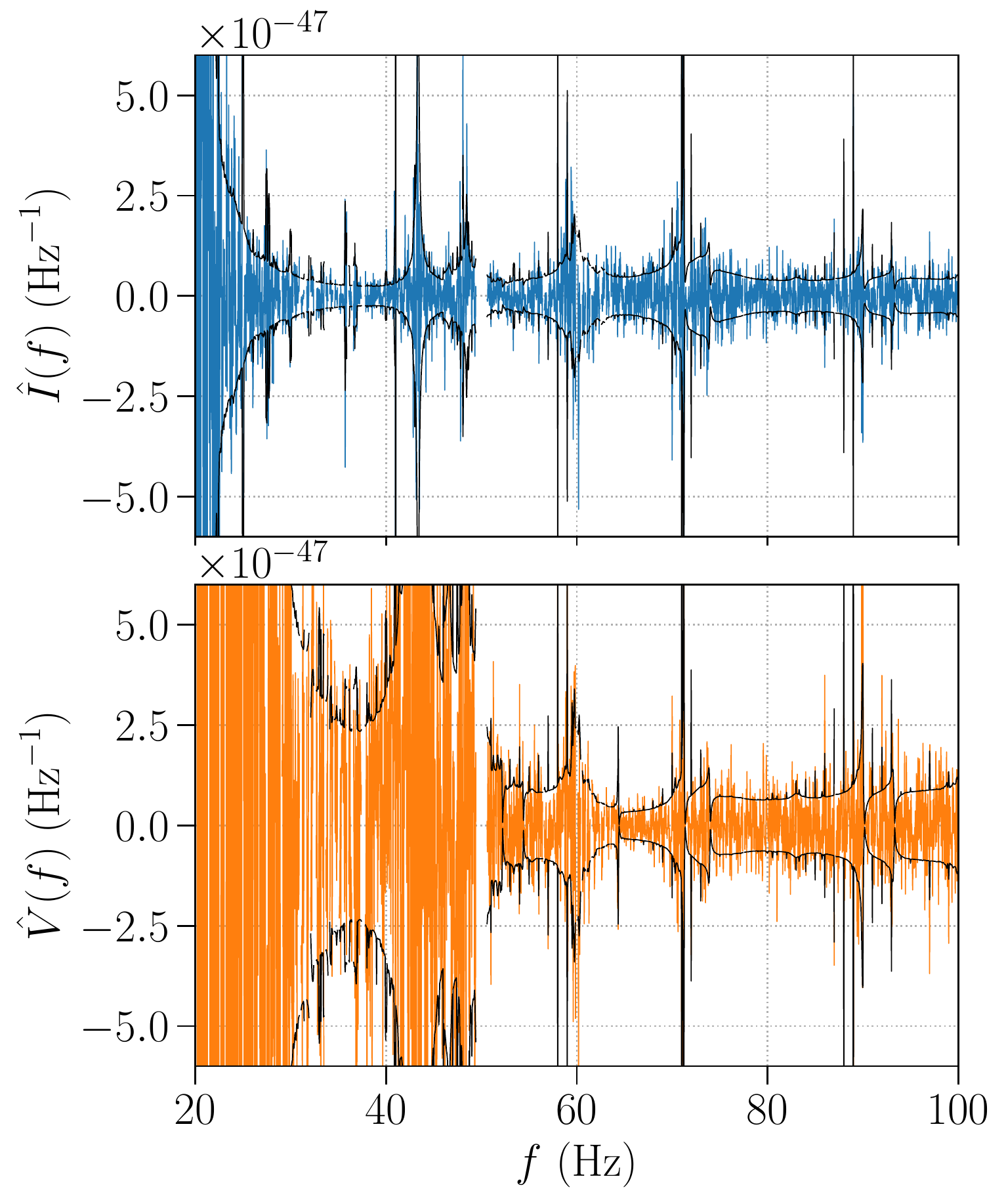}
    \caption{The estimation of Stokes' parameters $I(f)$ and $V(f)$ with LIGO-Virgo O1$\sim$ O3 runs. The black lines denote $1\sigma$ standard deviation which is equal to $\sqrt{\mathrm{Cov}_{ii}(f)}$.}
    \label{fig:IVestimation}
\end{figure}

The estimations of Stokes' parameters $I$ and $V$ in the frequency band $20\sim100$ Hz corresponding to the most sensitive band of LIGO-Virgo is plotted in \Fig{fig:IVestimation}. These results indicate that there is no evidence for the circularly polarized SGWB in the data of LIGO-Virgo's first three observing runs.  

\bigskip

{\it Power-law models. }
By adopting Bayesian inference technique according to Eq.~(\ref{likelihoodc}), we can place constraints on various GW spectra. In the LIGO-Virgo frequency band, for most theoretical models,  $\Omega_\text{gw}(f)$ can be approximated as a power-law: 
\begin{equation}
    \Omega_\text{gw}(f)=\Omega_\alpha\left(\frac{f}{f_\text{ref}}\right)^\alpha,
\end{equation}
where $\alpha$ is an index which is assumed to be a constant and the reference frequency $f_\text{ref}$ is taken to be 25 Hz in this letter.
In particular, cosmic string and slow roll inflation are well approximated by $\alpha=0$ in LIGO-Virgo frequency band \cite{PhysRevLett.85.3761,PhysRevD.55.R435,PhysRevD.50.1157}, CBCs produce a spectrum with $\alpha=2/3$ \cite{Regimbau:2011rp}, and $\alpha=3$ is a fiducial choice because it denotes a flat spectrum of $I(f)$ and some astrophysical processes like supernovae can produce such a kind of signal \cite{PhysRevD.73.104024}. 
In addition, we introduced a new parameter
\begin{equation}
    \Pi(f)=V(f)/I(f),
\end{equation}
which encodes the parity violation. If $\Pi(f)\neq 0$, the parity is violated. The range of $\Pi(f)$ is $[-1,1]$ in which the lower and upper bounds correspond to full left or right-handed polarizations, respectively.  For simplicity, $\Pi(f)$ is taken as a constant in our analysis.

\begin{figure}
    \centering
    \includegraphics[width=\columnwidth]{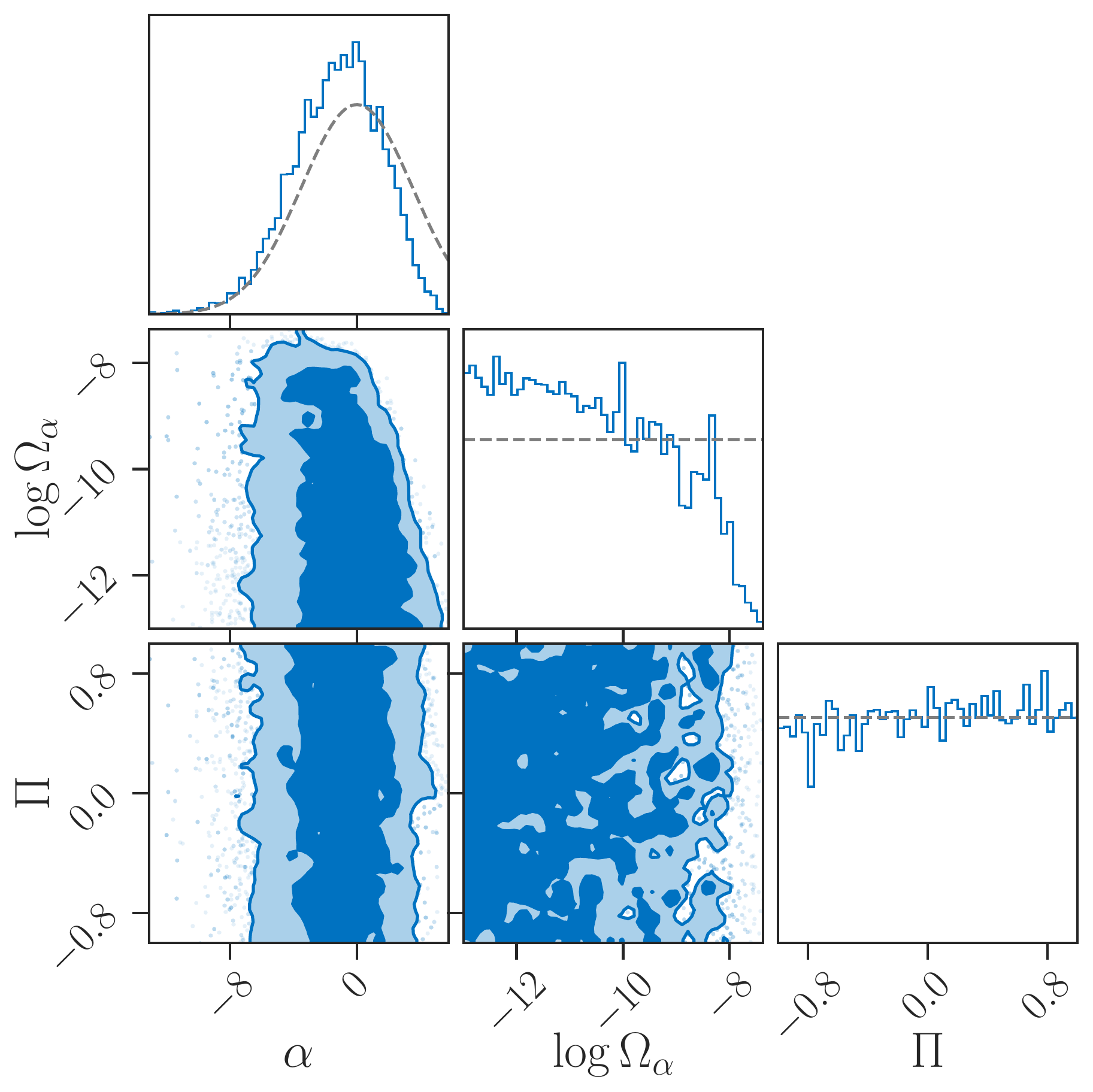}
    \caption{Posterior distributions of $\alpha$, $\log\Omega_\alpha$ and $\Pi$. The lower left panels are the 2D posterior density with $68\%$ and $95\%$ CL while the diagonal subplots show the marginalized 1D posteriors. The gray dotted lines denote priors adopted. }
    \label{fig:posterior}
\end{figure}

First of all, $\bm{\Theta}=(\Omega_\alpha,\alpha)$ and $\Pi$ are taken to be free parameters and $I(f)$ can be derived using \Eq{eq:IOmega}.  According to Bayes theorem, the posterior is given by 
\begin{equation}
    p(\bm{\Theta},\Pi|\bm{\hat{C}})\propto p(\bm{\hat{C}}|S[\Omega(f;\bm{\Theta}),\Pi])p(\bm{\Theta},\Pi).
\end{equation}
The ratio of evidence, so-called Bayes factor, is a factor measuring the relative possibility of hypotheses. Similar to \cite{LIGOScientific:2019vic,KAGRA:2021kbb}, we take a log-uniform prior for  $\Omega_\alpha$ and choose the lower bound to be $10^{-13}$. See \Table{tab:priors} for the priors for other parameters. Our results of the posterior distribution for the parameters are shown in \Fig{fig:posterior}. The Bayes factor is $\log\mathcal{B}=-0.2$ between signal and pure noise hypothesis, which indicates that there is no evidence for  claiming the existence of such signal. Besides, we have not found any significant restriction on the polarization parameter $\Pi$. At $95\%$ confidence level (CL), the 1D marginalized posterior of $\Omega_\alpha$ gives the upper limit of $5.3\times10^{-9}$ on the strength of SGWB at $25$ Hz.

\begin{table}[ht]
    \centering
    \begin{tabular}{p{0.25\columnwidth}<{\centering} p{0.55\columnwidth}<{\centering}}
        \Xhline{0.09em}
        Parameter & Prior  \\
        \hline
        $\Omega_\alpha$ & $\text{LogUniform}[10^{-13},\,10^{-7}]$ \\
        $\Pi$ & $\text{Uniform}[-1,\,1]$ \\
        $\alpha$ & $\text{Gaussian}(0,\,3.5^2)$ \\
        \Xhline{0.09em}
    \end{tabular}
    \caption{Prior distributions for free parameters in the analysis. We set a Gaussian prior with zero mean and standard deviation 3.5 for $\alpha$.}
    \label{tab:priors}
\end{table}%

\begin{figure*}[ht]
    \centering
    \includegraphics[width=0.33\textwidth]{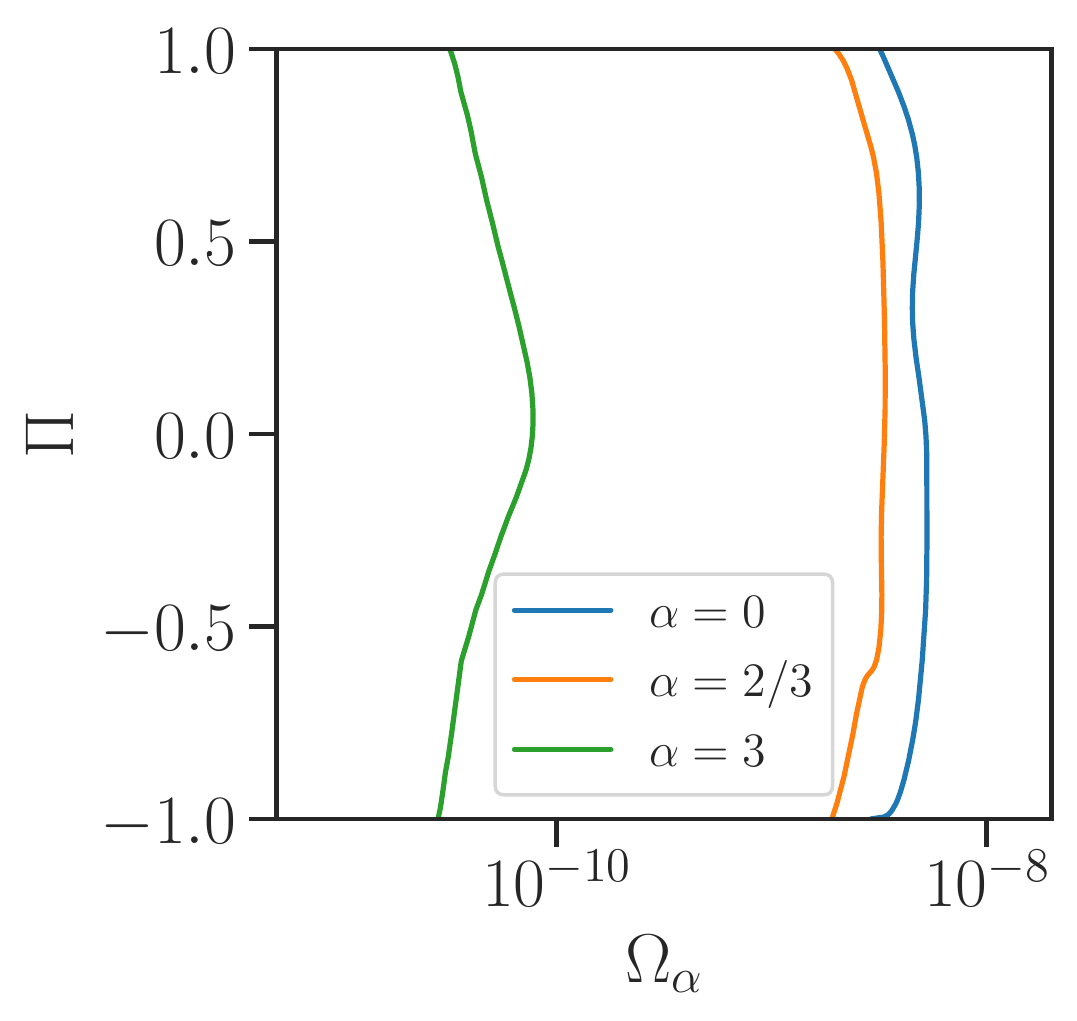}\;
    \includegraphics[width=0.315\textwidth]{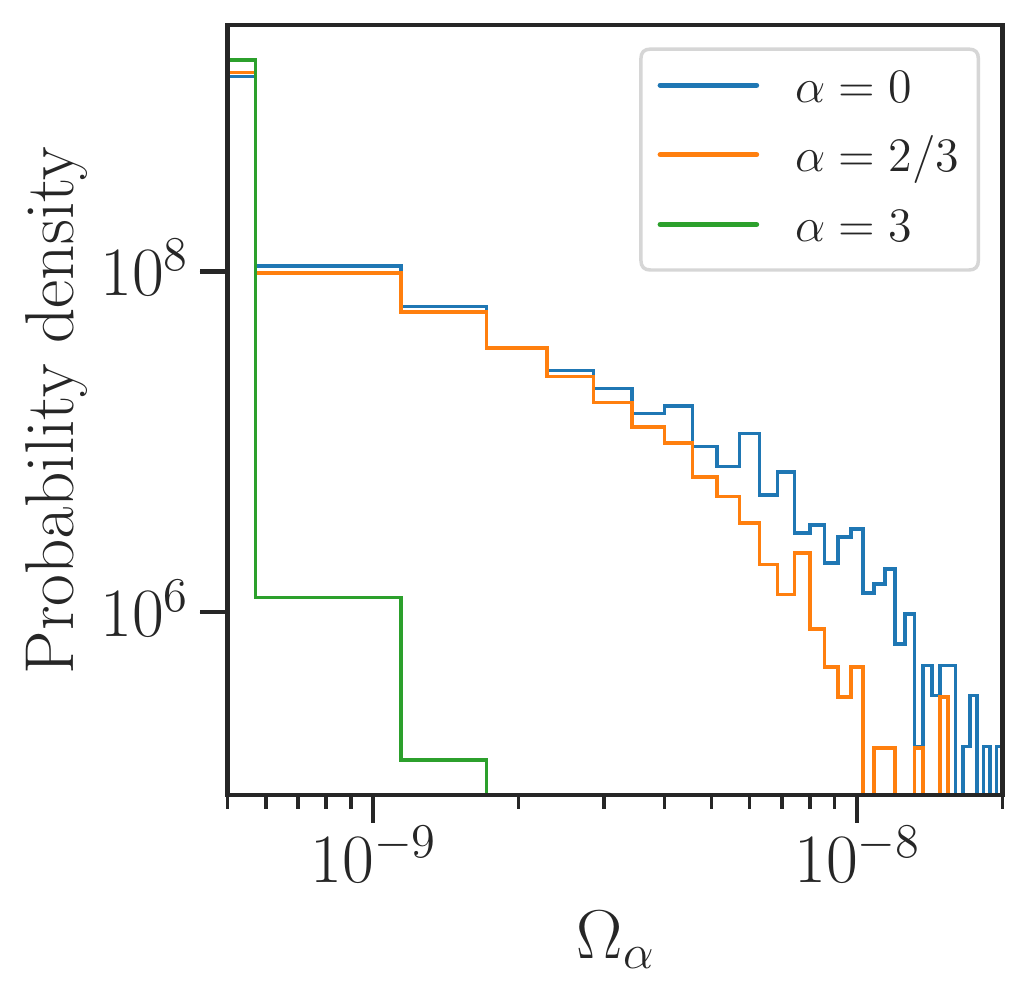}
    \includegraphics[width=0.326\textwidth]{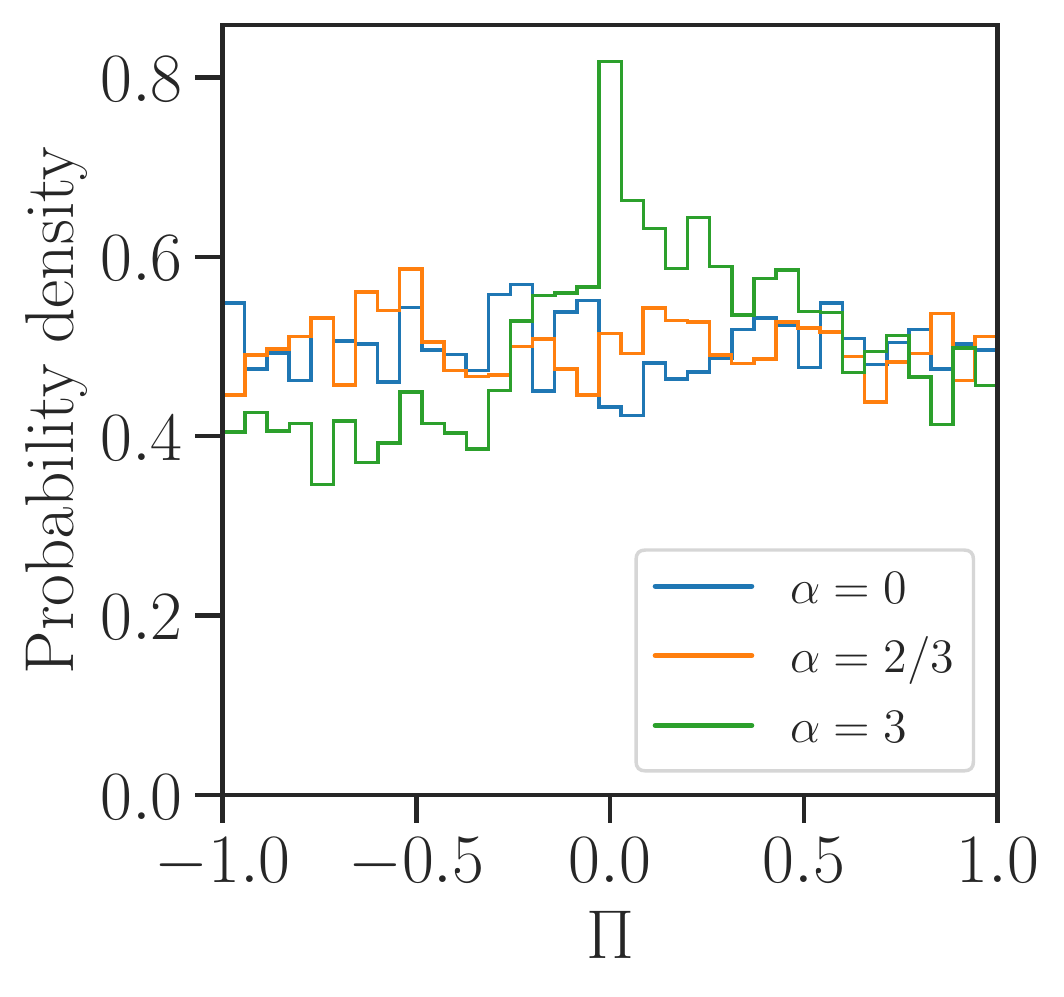}
    \caption{Left: Exclusion contours of the strength $\Omega_\alpha$ and polarization parameter $\Pi$ for three different power-law models at $95\%$ confidence levels. Center \& Right: Marginalized probability distributions of $\Omega_\alpha$ and $\Pi$.}
    \label{fig:contour1}
\end{figure*}

For some given theoretical models, $\alpha$ can be fixed. Therefore we also provide the constraints on the models with $\alpha=0, 2/3, 3$, respectively. Our results are given in \Fig{fig:contour1}. The left panel illustrates $95\%$ exclusion contours in $\Omega_\alpha$--$\Pi$ plane. The 1D probability distributions of $\Omega_\alpha$ and $\Pi$ are shown in center and right panels. 
To demonstrate how the constraints on SGWB strength  depend on the GW polarization, the upper limits of $\Omega_\alpha$ for the unpolarized $(\Pi=0)$ and fully polarized $(\Pi=\pm 1)$ cases are listed in \Table{tab:omega}. Due to the correlation contribution from $V(f)$, we are able to put tighter constraints on polarized SGWB, in particular  in the circumstance of $\alpha=3$. After marginalizing over the polarization degree, the upper limits for the power-law spectra with $\alpha=0, 2/3, 3$ are $3.6\times10^{-9}$, $2.3\times10^{-9}$ and $4.5\times10^{-11}$ at $95\%$ CL, respectively.

\begin{table}[ht]
    \centering
    \begin{tabular}{p{0.18\columnwidth}<{\centering} p{0.18\columnwidth}<{\centering} p{0.18\columnwidth}<{\centering} p{0.18\columnwidth}<{\centering} p{0.18\columnwidth}<{\centering}}
        \Xhline{0.09em}
         & $\Pi=-1$ & $\Pi=0$ & $\Pi=1$ & $\Pi$ Marg.\\
        \hline
        $\alpha=0$ & $3.7\times10^{-9}$  &  $4.0\times10^{-9}$ & $3.6\times10^{-9}$ & $3.6\times10^{-9}$ \\
        $\alpha=2/3$ & $1.7\times10^{-9}$  & $2.7\times10^{-9}$  & $1.7\times10^{-9}$ &  $2.3\times10^{-9}$ \\
        $\alpha=3$ & $1.1\times10^{-11}$  & $3.7\times10^{-10}$  & $1.6\times10^{-11}$ &  $4.5\times10^{-11}$ \\
        \Xhline{0.09em}
    \end{tabular}
    \caption{Upper limits of $\Omega_\alpha$ at $95\%$ CL for fixed values of $\alpha$ and $\Pi$. Besides, we also list the result after marginalizing $\Pi$.}
    \label{tab:omega}
\end{table}%

\bigskip

{\it Conclusions. }
In this work, the three independent baselines in the LIGO-Virgo network allow us to provide the first constraint on the circularly polarized isotropic SGWB by adopting maximum likelihood estimation and Bayesian statistics. Our results indicate that there is no evidence for the polarized SGWB in the data of Advanced LIGO-Virgo's first three observing runs.

By now, the ground-based detectors are not sensitive enough to claim a detection of SGWB. Moreover, the sensitivities of LIGO-Virgo pairs are worse than LIGO Hanford--Livingston pair. This further reduces the efficiency of detector networks and we are unable to obtain a significant constraint on the parity parameter $\Pi(f)$. But the extra baselines from LIGO-Virgo pairs do help to reduce the degeneracy between $\Omega_\alpha$ and $\Pi$. Therefore a lower upper limit for the strength of polarized SGWB compared to the unpolarized SGWB \cite{KAGRA:2021kbb} is obtained in our analysis.


\bigskip

{\it Acknowledgments. } 
We acknowledge the use of HPC Cluster of ITP-CAS. This work is supported by the National Key Research and Development Program of China Grant No.2020YFC2201502, grants from NSFC (grant No. 11975019, 11991052, 12047503), Key Research Program of Frontier Sciences, CAS, Grant NO. ZDBS-LY-7009, CAS Project for Young Scientists in Basic Research YSBR-006, the Key Research Program of the Chinese Academy of Sciences (Grant NO. XDPB15).

\bibliography{ref.bib}

\begin{thebibliography}{42}%
\makeatletter
\providecommand \@ifxundefined [1]{%
 \@ifx{#1\undefined}
}%
\providecommand \@ifnum [1]{%
 \ifnum #1\expandafter \@firstoftwo
 \else \expandafter \@secondoftwo
 \fi
}%
\providecommand \@ifx [1]{%
 \ifx #1\expandafter \@firstoftwo
 \else \expandafter \@secondoftwo
 \fi
}%
\providecommand \natexlab [1]{#1}%
\providecommand \enquote  [1]{``#1''}%
\providecommand \bibnamefont  [1]{#1}%
\providecommand \bibfnamefont [1]{#1}%
\providecommand \citenamefont [1]{#1}%
\providecommand \href@noop [0]{\@secondoftwo}%
\providecommand \href [0]{\begingroup \@sanitize@url \@href}%
\providecommand \@href[1]{\@@startlink{#1}\@@href}%
\providecommand \@@href[1]{\endgroup#1\@@endlink}%
\providecommand \@sanitize@url [0]{\catcode `\\12\catcode `\$12\catcode
  `\&12\catcode `\#12\catcode `\^12\catcode `\_12\catcode `\%12\relax}%
\providecommand \@@startlink[1]{}%
\providecommand \@@endlink[0]{}%
\providecommand \url  [0]{\begingroup\@sanitize@url \@url }%
\providecommand \@url [1]{\endgroup\@href {#1}{\urlprefix }}%
\providecommand \urlprefix  [0]{URL }%
\providecommand \Eprint [0]{\href }%
\providecommand \doibase [0]{http://dx.doi.org/}%
\providecommand \selectlanguage [0]{\@gobble}%
\providecommand \bibinfo  [0]{\@secondoftwo}%
\providecommand \bibfield  [0]{\@secondoftwo}%
\providecommand \translation [1]{[#1]}%
\providecommand \BibitemOpen [0]{}%
\providecommand \bibitemStop [0]{}%
\providecommand \bibitemNoStop [0]{.\EOS\space}%
\providecommand \EOS [0]{\spacefactor3000\relax}%
\providecommand \BibitemShut  [1]{\csname bibitem#1\endcsname}%
\let\auto@bib@innerbib\@empty
\bibitem [{\citenamefont {Rosado}(2011)}]{PhysRevD.84.084004}%
  \BibitemOpen
  \bibfield  {author} {\bibinfo {author} {\bibfnamefont {Pablo~A.}\
  \bibnamefont {Rosado}},\ }\bibfield  {title} {\enquote {\bibinfo {title}
  {Gravitational wave background from binary systems},}\ }\href {\doibase
  10.1103/PhysRevD.84.084004} {\bibfield  {journal} {\bibinfo  {journal} {Phys.
  Rev. D}\ }\textbf {\bibinfo {volume} {84}},\ \bibinfo {pages} {084004}
  (\bibinfo {year} {2011})}\BibitemShut {NoStop}%
\bibitem [{\citenamefont {Wu}\ \emph {et~al.}(2012)\citenamefont {Wu},
  \citenamefont {Mandic},\ and\ \citenamefont {Regimbau}}]{PhysRevD.85.104024}%
  \BibitemOpen
  \bibfield  {author} {\bibinfo {author} {\bibfnamefont {C.}~\bibnamefont
  {Wu}}, \bibinfo {author} {\bibfnamefont {V.}~\bibnamefont {Mandic}}, \ and\
  \bibinfo {author} {\bibfnamefont {T.}~\bibnamefont {Regimbau}},\ }\bibfield
  {title} {\enquote {\bibinfo {title} {Accessibility of the gravitational-wave
  background due to binary coalescences to second and third generation
  gravitational-wave detectors},}\ }\href {\doibase 10.1103/PhysRevD.85.104024}
  {\bibfield  {journal} {\bibinfo  {journal} {Phys. Rev. D}\ }\textbf {\bibinfo
  {volume} {85}},\ \bibinfo {pages} {104024} (\bibinfo {year}
  {2012})}\BibitemShut {NoStop}%
\bibitem [{\citenamefont {Zhu}\ \emph {et~al.}(2013)\citenamefont {Zhu},
  \citenamefont {Howell}, \citenamefont {Blair},\ and\ \citenamefont
  {Zhu}}]{Zhu:2012xw}%
  \BibitemOpen
  \bibfield  {author} {\bibinfo {author} {\bibfnamefont {Xing-Jiang}\
  \bibnamefont {Zhu}}, \bibinfo {author} {\bibfnamefont {Eric~J.}\ \bibnamefont
  {Howell}}, \bibinfo {author} {\bibfnamefont {David~G.}\ \bibnamefont
  {Blair}}, \ and\ \bibinfo {author} {\bibfnamefont {Zong-Hong}\ \bibnamefont
  {Zhu}},\ }\bibfield  {title} {\enquote {\bibinfo {title} {{On the
  gravitational wave background from compact binary coalescences in the band of
  ground-based interferometers}},}\ }\href {\doibase 10.1093/mnras/stt207}
  {\bibfield  {journal} {\bibinfo  {journal} {Mon. Not. Roy. Astron. Soc.}\
  }\textbf {\bibinfo {volume} {431}},\ \bibinfo {pages} {882--899} (\bibinfo
  {year} {2013})},\ \Eprint {http://arxiv.org/abs/1209.0595} {arXiv:1209.0595
  [gr-qc]} \BibitemShut {NoStop}%
\bibitem [{\citenamefont {Ferrari}\ \emph {et~al.}(1999)\citenamefont
  {Ferrari}, \citenamefont {Matarrese},\ and\ \citenamefont
  {Schneider}}]{Ferrari:1998jf}%
  \BibitemOpen
  \bibfield  {author} {\bibinfo {author} {\bibfnamefont {Valeria}\ \bibnamefont
  {Ferrari}}, \bibinfo {author} {\bibfnamefont {Sabino}\ \bibnamefont
  {Matarrese}}, \ and\ \bibinfo {author} {\bibfnamefont {Raffaella}\
  \bibnamefont {Schneider}},\ }\bibfield  {title} {\enquote {\bibinfo {title}
  {{Stochastic background of gravitational waves generated by a cosmological
  population of young, rapidly rotating neutron stars}},}\ }\href {\doibase
  10.1046/j.1365-8711.1999.02207.x} {\bibfield  {journal} {\bibinfo  {journal}
  {Mon. Not. Roy. Astron. Soc.}\ }\textbf {\bibinfo {volume} {303}},\ \bibinfo
  {pages} {258} (\bibinfo {year} {1999})},\ \Eprint
  {http://arxiv.org/abs/astro-ph/9806357} {arXiv:astro-ph/9806357} \BibitemShut
  {NoStop}%
\bibitem [{\citenamefont {Regimbau}\ and\ \citenamefont
  {de~Freitas~Pacheco}(2001)}]{Regimbau:2001kx}%
  \BibitemOpen
  \bibfield  {author} {\bibinfo {author} {\bibfnamefont {T.}~\bibnamefont
  {Regimbau}}\ and\ \bibinfo {author} {\bibfnamefont {Jose~A.}\ \bibnamefont
  {de~Freitas~Pacheco}},\ }\bibfield  {title} {\enquote {\bibinfo {title}
  {{Cosmic background of gravitational waves from rotating neutron stars}},}\
  }\href {\doibase 10.1051/0004-6361:20011005} {\bibfield  {journal} {\bibinfo
  {journal} {Astron. Astrophys.}\ }\textbf {\bibinfo {volume} {376}},\ \bibinfo
  {pages} {381} (\bibinfo {year} {2001})},\ \Eprint
  {http://arxiv.org/abs/astro-ph/0105260} {arXiv:astro-ph/0105260} \BibitemShut
  {NoStop}%
\bibitem [{\citenamefont {Lasky}\ \emph {et~al.}(2013)\citenamefont {Lasky},
  \citenamefont {Bennett},\ and\ \citenamefont {Melatos}}]{PhysRevD.87.063004}%
  \BibitemOpen
  \bibfield  {author} {\bibinfo {author} {\bibfnamefont {Paul~D.}\ \bibnamefont
  {Lasky}}, \bibinfo {author} {\bibfnamefont {Mark~F.}\ \bibnamefont
  {Bennett}}, \ and\ \bibinfo {author} {\bibfnamefont {Andrew}\ \bibnamefont
  {Melatos}},\ }\bibfield  {title} {\enquote {\bibinfo {title} {Stochastic
  gravitational wave background from hydrodynamic turbulence in differentially
  rotating neutron stars},}\ }\href {\doibase 10.1103/PhysRevD.87.063004}
  {\bibfield  {journal} {\bibinfo  {journal} {Phys. Rev. D}\ }\textbf {\bibinfo
  {volume} {87}},\ \bibinfo {pages} {063004} (\bibinfo {year}
  {2013})}\BibitemShut {NoStop}%
\bibitem [{\citenamefont {Crocker}\ \emph {et~al.}(2015)\citenamefont
  {Crocker}, \citenamefont {Mandic}, \citenamefont {Regimbau}, \citenamefont
  {Belczynski}, \citenamefont {Gladysz}, \citenamefont {Olive}, \citenamefont
  {Prestegard},\ and\ \citenamefont {Vangioni}}]{Crocker:2015taa}%
  \BibitemOpen
  \bibfield  {author} {\bibinfo {author} {\bibfnamefont {K.}~\bibnamefont
  {Crocker}}, \bibinfo {author} {\bibfnamefont {V.}~\bibnamefont {Mandic}},
  \bibinfo {author} {\bibfnamefont {T.}~\bibnamefont {Regimbau}}, \bibinfo
  {author} {\bibfnamefont {K.}~\bibnamefont {Belczynski}}, \bibinfo {author}
  {\bibfnamefont {W.}~\bibnamefont {Gladysz}}, \bibinfo {author} {\bibfnamefont
  {K.}~\bibnamefont {Olive}}, \bibinfo {author} {\bibfnamefont
  {T.}~\bibnamefont {Prestegard}}, \ and\ \bibinfo {author} {\bibfnamefont
  {E.}~\bibnamefont {Vangioni}},\ }\bibfield  {title} {\enquote {\bibinfo
  {title} {{Model of the stochastic gravitational-wave background due to core
  collapse to black holes}},}\ }\href {\doibase 10.1103/PhysRevD.92.063005}
  {\bibfield  {journal} {\bibinfo  {journal} {Phys. Rev. D}\ }\textbf {\bibinfo
  {volume} {92}},\ \bibinfo {pages} {063005} (\bibinfo {year} {2015})},\
  \Eprint {http://arxiv.org/abs/1506.02631} {arXiv:1506.02631 [gr-qc]}
  \BibitemShut {NoStop}%
\bibitem [{\citenamefont {Witten}(1984)}]{PhysRevD.30.272}%
  \BibitemOpen
  \bibfield  {author} {\bibinfo {author} {\bibfnamefont {Edward}\ \bibnamefont
  {Witten}},\ }\bibfield  {title} {\enquote {\bibinfo {title} {Cosmic
  separation of phases},}\ }\href {\doibase 10.1103/PhysRevD.30.272} {\bibfield
   {journal} {\bibinfo  {journal} {Phys. Rev. D}\ }\textbf {\bibinfo {volume}
  {30}},\ \bibinfo {pages} {272--285} (\bibinfo {year} {1984})}\BibitemShut
  {NoStop}%
\bibitem [{\citenamefont {Kosowsky}\ \emph {et~al.}(1992)\citenamefont
  {Kosowsky}, \citenamefont {Turner},\ and\ \citenamefont
  {Watkins}}]{Kosowsky:1992rz}%
  \BibitemOpen
  \bibfield  {author} {\bibinfo {author} {\bibfnamefont {Arthur}\ \bibnamefont
  {Kosowsky}}, \bibinfo {author} {\bibfnamefont {Michael~S.}\ \bibnamefont
  {Turner}}, \ and\ \bibinfo {author} {\bibfnamefont {Richard}\ \bibnamefont
  {Watkins}},\ }\bibfield  {title} {\enquote {\bibinfo {title} {{Gravitational
  waves from first order cosmological phase transitions}},}\ }\href {\doibase
  10.1103/PhysRevLett.69.2026} {\bibfield  {journal} {\bibinfo  {journal}
  {Phys. Rev. Lett.}\ }\textbf {\bibinfo {volume} {69}},\ \bibinfo {pages}
  {2026--2029} (\bibinfo {year} {1992})}\BibitemShut {NoStop}%
\bibitem [{\citenamefont {Dev}\ and\ \citenamefont
  {Mazumdar}(2016)}]{Dev:2016feu}%
  \BibitemOpen
  \bibfield  {author} {\bibinfo {author} {\bibfnamefont {P.~S.~Bhupal}\
  \bibnamefont {Dev}}\ and\ \bibinfo {author} {\bibfnamefont {A.}~\bibnamefont
  {Mazumdar}},\ }\bibfield  {title} {\enquote {\bibinfo {title} {{Probing the
  Scale of New Physics by Advanced LIGO/VIRGO}},}\ }\href {\doibase
  10.1103/PhysRevD.93.104001} {\bibfield  {journal} {\bibinfo  {journal} {Phys.
  Rev. D}\ }\textbf {\bibinfo {volume} {93}},\ \bibinfo {pages} {104001}
  (\bibinfo {year} {2016})},\ \Eprint {http://arxiv.org/abs/1602.04203}
  {arXiv:1602.04203 [hep-ph]} \BibitemShut {NoStop}%
\bibitem [{\citenamefont {Von~Harling}\ \emph {et~al.}(2020)\citenamefont
  {Von~Harling}, \citenamefont {Pomarol}, \citenamefont {Pujol\`as},\ and\
  \citenamefont {Rompineve}}]{VonHarling:2019rgb}%
  \BibitemOpen
  \bibfield  {author} {\bibinfo {author} {\bibfnamefont {Benedict}\
  \bibnamefont {Von~Harling}}, \bibinfo {author} {\bibfnamefont {Alex}\
  \bibnamefont {Pomarol}}, \bibinfo {author} {\bibfnamefont {Oriol}\
  \bibnamefont {Pujol\`as}}, \ and\ \bibinfo {author} {\bibfnamefont
  {Fabrizio}\ \bibnamefont {Rompineve}},\ }\bibfield  {title} {\enquote
  {\bibinfo {title} {{Peccei-Quinn Phase Transition at LIGO}},}\ }\href
  {\doibase 10.1007/JHEP04(2020)195} {\bibfield  {journal} {\bibinfo  {journal}
  {JHEP}\ }\textbf {\bibinfo {volume} {04}},\ \bibinfo {pages} {195} (\bibinfo
  {year} {2020})},\ \Eprint {http://arxiv.org/abs/1912.07587} {arXiv:1912.07587
  [hep-ph]} \BibitemShut {NoStop}%
\bibitem [{\citenamefont {Damour}\ and\ \citenamefont
  {Vilenkin}(2000)}]{PhysRevLett.85.3761}%
  \BibitemOpen
  \bibfield  {author} {\bibinfo {author} {\bibfnamefont {Thibault}\
  \bibnamefont {Damour}}\ and\ \bibinfo {author} {\bibfnamefont {Alexander}\
  \bibnamefont {Vilenkin}},\ }\bibfield  {title} {\enquote {\bibinfo {title}
  {Gravitational wave bursts from cosmic strings},}\ }\href {\doibase
  10.1103/PhysRevLett.85.3761} {\bibfield  {journal} {\bibinfo  {journal}
  {Phys. Rev. Lett.}\ }\textbf {\bibinfo {volume} {85}},\ \bibinfo {pages}
  {3761--3764} (\bibinfo {year} {2000})}\BibitemShut {NoStop}%
\bibitem [{\citenamefont {Sarangi}\ and\ \citenamefont
  {Tye}(2002)}]{Sarangi:2002yt}%
  \BibitemOpen
  \bibfield  {author} {\bibinfo {author} {\bibfnamefont {Saswat}\ \bibnamefont
  {Sarangi}}\ and\ \bibinfo {author} {\bibfnamefont {S.~H.~Henry}\ \bibnamefont
  {Tye}},\ }\bibfield  {title} {\enquote {\bibinfo {title} {{Cosmic string
  production towards the end of brane inflation}},}\ }\href {\doibase
  10.1016/S0370-2693(02)01824-5} {\bibfield  {journal} {\bibinfo  {journal}
  {Phys. Lett. B}\ }\textbf {\bibinfo {volume} {536}},\ \bibinfo {pages}
  {185--192} (\bibinfo {year} {2002})},\ \Eprint
  {http://arxiv.org/abs/hep-th/0204074} {arXiv:hep-th/0204074} \BibitemShut
  {NoStop}%
\bibitem [{\citenamefont {Siemens}\ \emph {et~al.}(2007)\citenamefont
  {Siemens}, \citenamefont {Mandic},\ and\ \citenamefont
  {Creighton}}]{PhysRevLett.98.111101}%
  \BibitemOpen
  \bibfield  {author} {\bibinfo {author} {\bibfnamefont {Xavier}\ \bibnamefont
  {Siemens}}, \bibinfo {author} {\bibfnamefont {Vuk}\ \bibnamefont {Mandic}}, \
  and\ \bibinfo {author} {\bibfnamefont {Jolien}\ \bibnamefont {Creighton}},\
  }\bibfield  {title} {\enquote {\bibinfo {title} {Gravitational-wave
  stochastic background from cosmic strings},}\ }\href {\doibase
  10.1103/PhysRevLett.98.111101} {\bibfield  {journal} {\bibinfo  {journal}
  {Phys. Rev. Lett.}\ }\textbf {\bibinfo {volume} {98}},\ \bibinfo {pages}
  {111101} (\bibinfo {year} {2007})}\BibitemShut {NoStop}%
\bibitem [{\citenamefont {Abbott}\ \emph {et~al.}(2018)\citenamefont {Abbott}
  \emph {et~al.}}]{LIGOScientific:2017ikf}%
  \BibitemOpen
  \bibfield  {author} {\bibinfo {author} {\bibfnamefont {B.~P.}\ \bibnamefont
  {Abbott}} \emph {et~al.} (\bibinfo {collaboration} {LIGO Scientific,
  Virgo}),\ }\bibfield  {title} {\enquote {\bibinfo {title} {{Constraints on
  cosmic strings using data from the first Advanced LIGO observing run}},}\
  }\href {\doibase 10.1103/PhysRevD.97.102002} {\bibfield  {journal} {\bibinfo
  {journal} {Phys. Rev. D}\ }\textbf {\bibinfo {volume} {97}},\ \bibinfo
  {pages} {102002} (\bibinfo {year} {2018})},\ \Eprint
  {http://arxiv.org/abs/1712.01168} {arXiv:1712.01168 [gr-qc]} \BibitemShut
  {NoStop}%
\bibitem [{\citenamefont {Turner}(1997)}]{PhysRevD.55.R435}%
  \BibitemOpen
  \bibfield  {author} {\bibinfo {author} {\bibfnamefont {Michael~S.}\
  \bibnamefont {Turner}},\ }\bibfield  {title} {\enquote {\bibinfo {title}
  {Detectability of inflation-produced gravitational waves},}\ }\href {\doibase
  10.1103/PhysRevD.55.R435} {\bibfield  {journal} {\bibinfo  {journal} {Phys.
  Rev. D}\ }\textbf {\bibinfo {volume} {55}},\ \bibinfo {pages} {R435--R439}
  (\bibinfo {year} {1997})}\BibitemShut {NoStop}%
\bibitem [{\citenamefont {Guzzetti}\ \emph {et~al.}(2016)\citenamefont
  {Guzzetti}, \citenamefont {Bartolo}, \citenamefont {Liguori},\ and\
  \citenamefont {Matarrese}}]{Guzzetti:2016mkm}%
  \BibitemOpen
  \bibfield  {author} {\bibinfo {author} {\bibfnamefont {M.~C.}\ \bibnamefont
  {Guzzetti}}, \bibinfo {author} {\bibfnamefont {N.}~\bibnamefont {Bartolo}},
  \bibinfo {author} {\bibfnamefont {M.}~\bibnamefont {Liguori}}, \ and\
  \bibinfo {author} {\bibfnamefont {S.}~\bibnamefont {Matarrese}},\ }\bibfield
  {title} {\enquote {\bibinfo {title} {{Gravitational waves from inflation}},}\
  }\href {\doibase 10.1393/ncr/i2016-10127-1} {\bibfield  {journal} {\bibinfo
  {journal} {Riv. Nuovo Cim.}\ }\textbf {\bibinfo {volume} {39}},\ \bibinfo
  {pages} {399--495} (\bibinfo {year} {2016})},\ \Eprint
  {http://arxiv.org/abs/1605.01615} {arXiv:1605.01615 [astro-ph.CO]}
  \BibitemShut {NoStop}%
\bibitem [{\citenamefont {Aasi}\ \emph {et~al.}(2015)\citenamefont {Aasi},
  \citenamefont {Abbott}, \citenamefont {Abbott}, \citenamefont {Abbott},
  \citenamefont {Abernathy}, \citenamefont {Ackley}, \citenamefont {Adams},
  \citenamefont {Adams}, \citenamefont {Addesso}, \citenamefont {Adhikari},
  \citenamefont {Adya}, \citenamefont {Affeldt}, \citenamefont {Aggarwal},
  \citenamefont {Aguiar}, \citenamefont {Ain}, \citenamefont {Ajith},
  \citenamefont {Alemic}, \citenamefont {Allen}, \citenamefont {Amariutei},\
  and\ \citenamefont {Zweizig}}]{AdvancedLIGO2015}%
  \BibitemOpen
  \bibfield  {author} {\bibinfo {author} {\bibfnamefont {J}~\bibnamefont
  {Aasi}}, \bibinfo {author} {\bibfnamefont {B}~\bibnamefont {Abbott}},
  \bibinfo {author} {\bibfnamefont {R}~\bibnamefont {Abbott}}, \bibinfo
  {author} {\bibfnamefont {Temeka}\ \bibnamefont {Abbott}}, \bibinfo {author}
  {\bibfnamefont {Matthew}\ \bibnamefont {Abernathy}}, \bibinfo {author}
  {\bibfnamefont {K}~\bibnamefont {Ackley}}, \bibinfo {author} {\bibfnamefont
  {C}~\bibnamefont {Adams}}, \bibinfo {author} {\bibfnamefont {Teneisha}\
  \bibnamefont {Adams}}, \bibinfo {author} {\bibfnamefont {Paolo}\ \bibnamefont
  {Addesso}}, \bibinfo {author} {\bibfnamefont {R}~\bibnamefont {Adhikari}},
  \bibinfo {author} {\bibfnamefont {Vaishali}\ \bibnamefont {Adya}}, \bibinfo
  {author} {\bibfnamefont {C}~\bibnamefont {Affeldt}}, \bibinfo {author}
  {\bibfnamefont {Nishu}\ \bibnamefont {Aggarwal}}, \bibinfo {author}
  {\bibfnamefont {Odylio}\ \bibnamefont {Aguiar}}, \bibinfo {author}
  {\bibfnamefont {Anirban}\ \bibnamefont {Ain}}, \bibinfo {author}
  {\bibfnamefont {P}~\bibnamefont {Ajith}}, \bibinfo {author} {\bibfnamefont
  {A}~\bibnamefont {Alemic}}, \bibinfo {author} {\bibfnamefont {Bruce}\
  \bibnamefont {Allen}}, \bibinfo {author} {\bibfnamefont {D}~\bibnamefont
  {Amariutei}}, \ and\ \bibinfo {author} {\bibfnamefont {John}\ \bibnamefont
  {Zweizig}},\ }\bibfield  {title} {\enquote {\bibinfo {title} {Advanced
  ligo},}\ }\href {\doibase 10.1088/0264-9381/32/7/074001} {\bibfield
  {journal} {\bibinfo  {journal} {Classical and Quantum Gravity}\ }\textbf
  {\bibinfo {volume} {32}} (\bibinfo {year} {2015}),\
  10.1088/0264-9381/32/7/074001}\BibitemShut {NoStop}%
\bibitem [{\citenamefont {Acernese}\ \emph {et~al.}(2015)\citenamefont
  {Acernese} \emph {et~al.}}]{VIRGO:2014yos}%
  \BibitemOpen
  \bibfield  {author} {\bibinfo {author} {\bibfnamefont {F.}~\bibnamefont
  {Acernese}} \emph {et~al.} (\bibinfo {collaboration} {VIRGO}),\ }\bibfield
  {title} {\enquote {\bibinfo {title} {{Advanced Virgo: a second-generation
  interferometric gravitational wave detector}},}\ }\href {\doibase
  10.1088/0264-9381/32/2/024001} {\bibfield  {journal} {\bibinfo  {journal}
  {Class. Quant. Grav.}\ }\textbf {\bibinfo {volume} {32}},\ \bibinfo {pages}
  {024001} (\bibinfo {year} {2015})},\ \Eprint {http://arxiv.org/abs/1408.3978}
  {arXiv:1408.3978 [gr-qc]} \BibitemShut {NoStop}%
\bibitem [{\citenamefont {Abbott}\ \emph
  {et~al.}(2021{\natexlab{a}})\citenamefont {Abbott} \emph
  {et~al.}}]{KAGRA:2021kbb}%
  \BibitemOpen
  \bibfield  {author} {\bibinfo {author} {\bibfnamefont {R.}~\bibnamefont
  {Abbott}} \emph {et~al.} (\bibinfo {collaboration} {KAGRA, Virgo, LIGO
  Scientific}),\ }\bibfield  {title} {\enquote {\bibinfo {title} {{Upper limits
  on the isotropic gravitational-wave background from Advanced LIGO and
  Advanced Virgo\textquoteright{}s third observing run}},}\ }\href {\doibase
  10.1103/PhysRevD.104.022004} {\bibfield  {journal} {\bibinfo  {journal}
  {Phys. Rev. D}\ }\textbf {\bibinfo {volume} {104}},\ \bibinfo {pages}
  {022004} (\bibinfo {year} {2021}{\natexlab{a}})},\ \Eprint
  {http://arxiv.org/abs/2101.12130} {arXiv:2101.12130 [gr-qc]} \BibitemShut
  {NoStop}%
\bibitem [{\citenamefont {Abbott}\ \emph
  {et~al.}(2021{\natexlab{b}})\citenamefont {Abbott} \emph
  {et~al.}}]{KAGRA:2021mth}%
  \BibitemOpen
  \bibfield  {author} {\bibinfo {author} {\bibfnamefont {R.}~\bibnamefont
  {Abbott}} \emph {et~al.} (\bibinfo {collaboration} {KAGRA, Virgo, LIGO
  Scientific}),\ }\bibfield  {title} {\enquote {\bibinfo {title} {{Search for
  anisotropic gravitational-wave backgrounds using data from Advanced LIGO and
  Advanced Virgo\textquoteright{}s first three observing runs}},}\ }\href
  {\doibase 10.1103/PhysRevD.104.022005} {\bibfield  {journal} {\bibinfo
  {journal} {Phys. Rev. D}\ }\textbf {\bibinfo {volume} {104}},\ \bibinfo
  {pages} {022005} (\bibinfo {year} {2021}{\natexlab{b}})},\ \Eprint
  {http://arxiv.org/abs/2103.08520} {arXiv:2103.08520 [gr-qc]} \BibitemShut
  {NoStop}%
\bibitem [{\citenamefont {Abbott}\ \emph {et~al.}(2022)\citenamefont {Abbott}
  \emph {et~al.}}]{LIGOScientific:2021qeg}%
  \BibitemOpen
  \bibfield  {author} {\bibinfo {author} {\bibfnamefont {R.}~\bibnamefont
  {Abbott}} \emph {et~al.} (\bibinfo {collaboration} {LIGO Scientific, Virgo,
  KAGRA}),\ }\bibfield  {title} {\enquote {\bibinfo {title} {{All-sky,
  all-frequency directional search for persistent gravitational waves from
  Advanced LIGO\textquoteright{}s and Advanced Virgo\textquoteright{}s first
  three observing runs}},}\ }\href {\doibase 10.1103/PhysRevD.105.122001}
  {\bibfield  {journal} {\bibinfo  {journal} {Phys. Rev. D}\ }\textbf {\bibinfo
  {volume} {105}},\ \bibinfo {pages} {122001} (\bibinfo {year} {2022})},\
  \Eprint {http://arxiv.org/abs/2110.09834} {arXiv:2110.09834 [gr-qc]}
  \BibitemShut {NoStop}%
\bibitem [{\citenamefont {Kahniashvili}\ \emph {et~al.}(2005)\citenamefont
  {Kahniashvili}, \citenamefont {Gogoberidze},\ and\ \citenamefont
  {Ratra}}]{Kahniashvili:2005qi}%
  \BibitemOpen
  \bibfield  {author} {\bibinfo {author} {\bibfnamefont {Tina}\ \bibnamefont
  {Kahniashvili}}, \bibinfo {author} {\bibfnamefont {Grigol}\ \bibnamefont
  {Gogoberidze}}, \ and\ \bibinfo {author} {\bibfnamefont {Bharat}\
  \bibnamefont {Ratra}},\ }\bibfield  {title} {\enquote {\bibinfo {title}
  {{Polarized cosmological gravitational waves from primordial helical
  turbulence}},}\ }\href {\doibase 10.1103/PhysRevLett.95.151301} {\bibfield
  {journal} {\bibinfo  {journal} {Phys. Rev. Lett.}\ }\textbf {\bibinfo
  {volume} {95}},\ \bibinfo {pages} {151301} (\bibinfo {year} {2005})},\
  \Eprint {http://arxiv.org/abs/astro-ph/0505628} {arXiv:astro-ph/0505628}
  \BibitemShut {NoStop}%
\bibitem [{\citenamefont {Alexander}\ and\ \citenamefont
  {Yunes}(2009)}]{Alexander:2009tp}%
  \BibitemOpen
  \bibfield  {author} {\bibinfo {author} {\bibfnamefont {Stephon}\ \bibnamefont
  {Alexander}}\ and\ \bibinfo {author} {\bibfnamefont {Nicolas}\ \bibnamefont
  {Yunes}},\ }\bibfield  {title} {\enquote {\bibinfo {title} {{Chern-Simons
  Modified General Relativity}},}\ }\href {\doibase
  10.1016/j.physrep.2009.07.002} {\bibfield  {journal} {\bibinfo  {journal}
  {Phys. Rept.}\ }\textbf {\bibinfo {volume} {480}},\ \bibinfo {pages} {1--55}
  (\bibinfo {year} {2009})},\ \Eprint {http://arxiv.org/abs/0907.2562}
  {arXiv:0907.2562 [hep-th]} \BibitemShut {NoStop}%
\bibitem [{\citenamefont {Lyth}\ \emph {et~al.}(2005)\citenamefont {Lyth},
  \citenamefont {Quimbay},\ and\ \citenamefont {Rodriguez}}]{Lyth:2005jf}%
  \BibitemOpen
  \bibfield  {author} {\bibinfo {author} {\bibfnamefont {David~H.}\
  \bibnamefont {Lyth}}, \bibinfo {author} {\bibfnamefont {Carlos}\ \bibnamefont
  {Quimbay}}, \ and\ \bibinfo {author} {\bibfnamefont {Yeinzon}\ \bibnamefont
  {Rodriguez}},\ }\bibfield  {title} {\enquote {\bibinfo {title} {{Leptogenesis
  and tensor polarisation from a gravitational Chern-Simons term}},}\ }\href
  {\doibase 10.1088/1126-6708/2005/03/016} {\bibfield  {journal} {\bibinfo
  {journal} {JHEP}\ }\textbf {\bibinfo {volume} {03}},\ \bibinfo {pages} {016}
  (\bibinfo {year} {2005})},\ \Eprint {http://arxiv.org/abs/hep-th/0501153}
  {arXiv:hep-th/0501153} \BibitemShut {NoStop}%
\bibitem [{\citenamefont {Satoh}\ \emph {et~al.}(2008)\citenamefont {Satoh},
  \citenamefont {Kanno},\ and\ \citenamefont {Soda}}]{Satoh:2007gn}%
  \BibitemOpen
  \bibfield  {author} {\bibinfo {author} {\bibfnamefont {Masaki}\ \bibnamefont
  {Satoh}}, \bibinfo {author} {\bibfnamefont {Sugumi}\ \bibnamefont {Kanno}}, \
  and\ \bibinfo {author} {\bibfnamefont {Jiro}\ \bibnamefont {Soda}},\
  }\bibfield  {title} {\enquote {\bibinfo {title} {{Circular Polarization of
  Primordial Gravitational Waves in String-inspired Inflationary Cosmology}},}\
  }\href {\doibase 10.1103/PhysRevD.77.023526} {\bibfield  {journal} {\bibinfo
  {journal} {Phys. Rev. D}\ }\textbf {\bibinfo {volume} {77}},\ \bibinfo
  {pages} {023526} (\bibinfo {year} {2008})},\ \Eprint
  {http://arxiv.org/abs/0706.3585} {arXiv:0706.3585 [astro-ph]} \BibitemShut
  {NoStop}%
\bibitem [{\citenamefont {Saito}\ \emph {et~al.}(2007)\citenamefont {Saito},
  \citenamefont {Ichiki},\ and\ \citenamefont {Taruya}}]{Saito:2007kt}%
  \BibitemOpen
  \bibfield  {author} {\bibinfo {author} {\bibfnamefont {Shun}\ \bibnamefont
  {Saito}}, \bibinfo {author} {\bibfnamefont {Kiyotomo}\ \bibnamefont
  {Ichiki}}, \ and\ \bibinfo {author} {\bibfnamefont {Atsushi}\ \bibnamefont
  {Taruya}},\ }\bibfield  {title} {\enquote {\bibinfo {title} {{Probing
  polarization states of primordial gravitational waves with CMB
  anisotropies}},}\ }\href {\doibase 10.1088/1475-7516/2007/09/002} {\bibfield
  {journal} {\bibinfo  {journal} {JCAP}\ }\textbf {\bibinfo {volume} {09}},\
  \bibinfo {pages} {002} (\bibinfo {year} {2007})},\ \Eprint
  {http://arxiv.org/abs/0705.3701} {arXiv:0705.3701 [astro-ph]} \BibitemShut
  {NoStop}%
\bibitem [{\citenamefont {Horava}(2009)}]{Horava:2009uw}%
  \BibitemOpen
  \bibfield  {author} {\bibinfo {author} {\bibfnamefont {Petr}\ \bibnamefont
  {Horava}},\ }\bibfield  {title} {\enquote {\bibinfo {title} {{Quantum Gravity
  at a Lifshitz Point}},}\ }\href {\doibase 10.1103/PhysRevD.79.084008}
  {\bibfield  {journal} {\bibinfo  {journal} {Phys. Rev. D}\ }\textbf {\bibinfo
  {volume} {79}},\ \bibinfo {pages} {084008} (\bibinfo {year} {2009})},\
  \Eprint {http://arxiv.org/abs/0901.3775} {arXiv:0901.3775 [hep-th]}
  \BibitemShut {NoStop}%
\bibitem [{\citenamefont {Takahashi}\ and\ \citenamefont
  {Soda}(2009)}]{PhysRevLett.102.231301}%
  \BibitemOpen
  \bibfield  {author} {\bibinfo {author} {\bibfnamefont {Tomohiro}\
  \bibnamefont {Takahashi}}\ and\ \bibinfo {author} {\bibfnamefont {Jiro}\
  \bibnamefont {Soda}},\ }\bibfield  {title} {\enquote {\bibinfo {title}
  {Chiral primordial gravitational waves from a lifshitz point},}\ }\href
  {\doibase 10.1103/PhysRevLett.102.231301} {\bibfield  {journal} {\bibinfo
  {journal} {Phys. Rev. Lett.}\ }\textbf {\bibinfo {volume} {102}},\ \bibinfo
  {pages} {231301} (\bibinfo {year} {2009})}\BibitemShut {NoStop}%
\bibitem [{\citenamefont {Crisostomi}\ \emph {et~al.}(2018)\citenamefont
  {Crisostomi}, \citenamefont {Noui}, \citenamefont {Charmousis},\ and\
  \citenamefont {Langlois}}]{Crisostomi:2017ugk}%
  \BibitemOpen
  \bibfield  {author} {\bibinfo {author} {\bibfnamefont {Marco}\ \bibnamefont
  {Crisostomi}}, \bibinfo {author} {\bibfnamefont {Karim}\ \bibnamefont
  {Noui}}, \bibinfo {author} {\bibfnamefont {Christos}\ \bibnamefont
  {Charmousis}}, \ and\ \bibinfo {author} {\bibfnamefont {David}\ \bibnamefont
  {Langlois}},\ }\bibfield  {title} {\enquote {\bibinfo {title} {{Beyond
  Lovelock gravity: Higher derivative metric theories}},}\ }\href {\doibase
  10.1103/PhysRevD.97.044034} {\bibfield  {journal} {\bibinfo  {journal} {Phys.
  Rev. D}\ }\textbf {\bibinfo {volume} {97}},\ \bibinfo {pages} {044034}
  (\bibinfo {year} {2018})},\ \Eprint {http://arxiv.org/abs/1710.04531}
  {arXiv:1710.04531 [hep-th]} \BibitemShut {NoStop}%
\bibitem [{\citenamefont {Zhao}\ \emph {et~al.}(2020)\citenamefont {Zhao},
  \citenamefont {Zhu}, \citenamefont {Qiao},\ and\ \citenamefont
  {Wang}}]{Zhao:2019xmm}%
  \BibitemOpen
  \bibfield  {author} {\bibinfo {author} {\bibfnamefont {Wen}\ \bibnamefont
  {Zhao}}, \bibinfo {author} {\bibfnamefont {Tao}\ \bibnamefont {Zhu}},
  \bibinfo {author} {\bibfnamefont {Jin}\ \bibnamefont {Qiao}}, \ and\ \bibinfo
  {author} {\bibfnamefont {Anzhong}\ \bibnamefont {Wang}},\ }\bibfield  {title}
  {\enquote {\bibinfo {title} {{Waveform of gravitational waves in the general
  parity-violating gravities}},}\ }\href {\doibase 10.1103/PhysRevD.101.024002}
  {\bibfield  {journal} {\bibinfo  {journal} {Phys. Rev. D}\ }\textbf {\bibinfo
  {volume} {101}},\ \bibinfo {pages} {024002} (\bibinfo {year} {2020})},\
  \Eprint {http://arxiv.org/abs/1909.10887} {arXiv:1909.10887 [gr-qc]}
  \BibitemShut {NoStop}%
\bibitem [{\citenamefont {Seto}\ and\ \citenamefont
  {Taruya}(2007)}]{Seto:2007tn}%
  \BibitemOpen
  \bibfield  {author} {\bibinfo {author} {\bibfnamefont {Naoki}\ \bibnamefont
  {Seto}}\ and\ \bibinfo {author} {\bibfnamefont {Atsushi}\ \bibnamefont
  {Taruya}},\ }\bibfield  {title} {\enquote {\bibinfo {title} {{Measuring a
  Parity Violation Signature in the Early Universe via Ground-based Laser
  Interferometers}},}\ }\href {\doibase 10.1103/PhysRevLett.99.121101}
  {\bibfield  {journal} {\bibinfo  {journal} {Phys. Rev. Lett.}\ }\textbf
  {\bibinfo {volume} {99}},\ \bibinfo {pages} {121101} (\bibinfo {year}
  {2007})},\ \Eprint {http://arxiv.org/abs/0707.0535} {arXiv:0707.0535
  [astro-ph]} \BibitemShut {NoStop}%
\bibitem [{\citenamefont {Seto}\ and\ \citenamefont
  {Taruya}(2008)}]{Seto:2008sr}%
  \BibitemOpen
  \bibfield  {author} {\bibinfo {author} {\bibfnamefont {Naoki}\ \bibnamefont
  {Seto}}\ and\ \bibinfo {author} {\bibfnamefont {Atsushi}\ \bibnamefont
  {Taruya}},\ }\bibfield  {title} {\enquote {\bibinfo {title} {{Polarization
  analysis of gravitational-wave backgrounds from the correlation signals of
  ground-based interferometers: Measuring a circular-polarization mode}},}\
  }\href {\doibase 10.1103/PhysRevD.77.103001} {\bibfield  {journal} {\bibinfo
  {journal} {Phys. Rev. D}\ }\textbf {\bibinfo {volume} {77}},\ \bibinfo
  {pages} {103001} (\bibinfo {year} {2008})},\ \Eprint
  {http://arxiv.org/abs/0801.4185} {arXiv:0801.4185 [astro-ph]} \BibitemShut
  {NoStop}%
\bibitem [{\citenamefont {Crowder}\ \emph {et~al.}(2013)\citenamefont
  {Crowder}, \citenamefont {Namba}, \citenamefont {Mandic}, \citenamefont
  {Mukohyama},\ and\ \citenamefont {Peloso}}]{Crowder:2012ik}%
  \BibitemOpen
  \bibfield  {author} {\bibinfo {author} {\bibfnamefont {S.~G.}\ \bibnamefont
  {Crowder}}, \bibinfo {author} {\bibfnamefont {R.}~\bibnamefont {Namba}},
  \bibinfo {author} {\bibfnamefont {V.}~\bibnamefont {Mandic}}, \bibinfo
  {author} {\bibfnamefont {S.}~\bibnamefont {Mukohyama}}, \ and\ \bibinfo
  {author} {\bibfnamefont {M.}~\bibnamefont {Peloso}},\ }\bibfield  {title}
  {\enquote {\bibinfo {title} {{Measurement of Parity Violation in the Early
  Universe using Gravitational-wave Detectors}},}\ }\href {\doibase
  10.1016/j.physletb.2013.08.077} {\bibfield  {journal} {\bibinfo  {journal}
  {Phys. Lett. B}\ }\textbf {\bibinfo {volume} {726}},\ \bibinfo {pages}
  {66--71} (\bibinfo {year} {2013})},\ \Eprint {http://arxiv.org/abs/1212.4165}
  {arXiv:1212.4165 [astro-ph.CO]} \BibitemShut {NoStop}%
\bibitem [{\citenamefont {Allen}\ and\ \citenamefont
  {Romano}(1999)}]{Allen:1997ad}%
  \BibitemOpen
  \bibfield  {author} {\bibinfo {author} {\bibfnamefont {Bruce}\ \bibnamefont
  {Allen}}\ and\ \bibinfo {author} {\bibfnamefont {Joseph~D.}\ \bibnamefont
  {Romano}},\ }\bibfield  {title} {\enquote {\bibinfo {title} {{Detecting a
  stochastic background of gravitational radiation: Signal processing
  strategies and sensitivities}},}\ }\href {\doibase
  10.1103/PhysRevD.59.102001} {\bibfield  {journal} {\bibinfo  {journal} {Phys.
  Rev. D}\ }\textbf {\bibinfo {volume} {59}},\ \bibinfo {pages} {102001}
  (\bibinfo {year} {1999})},\ \Eprint {http://arxiv.org/abs/gr-qc/9710117}
  {arXiv:gr-qc/9710117} \BibitemShut {NoStop}%
\bibitem [{\citenamefont {Abbott}\ \emph {et~al.}()\citenamefont {Abbott} \emph
  {et~al.}}]{LVK:IsoSGWBdata}%
  \BibitemOpen
  \bibfield  {author} {\bibinfo {author} {\bibfnamefont {R.}~\bibnamefont
  {Abbott}} \emph {et~al.} (\bibinfo {collaboration} {LIGO Scientific, Virgo,
  KAGRA}),\ }\href@noop {} {}\bibinfo {howpublished}
  {\url{https://dcc.ligo.org/G2001287/public}}\BibitemShut {NoStop}%
\bibitem [{\citenamefont {{Thrane}}\ \emph {et~al.}(2009)\citenamefont
  {{Thrane}}, \citenamefont {{Ballmer}}, \citenamefont {{Romano}},
  \citenamefont {{Mitra}}, \citenamefont {{Talukder}}, \citenamefont {{Bose}},\
  and\ \citenamefont {{Mandic}}}]{2009PhRvD..80l2002T}%
  \BibitemOpen
  \bibfield  {author} {\bibinfo {author} {\bibfnamefont {Eric}\ \bibnamefont
  {{Thrane}}}, \bibinfo {author} {\bibfnamefont {Stefan}\ \bibnamefont
  {{Ballmer}}}, \bibinfo {author} {\bibfnamefont {Joseph~D.}\ \bibnamefont
  {{Romano}}}, \bibinfo {author} {\bibfnamefont {Sanjit}\ \bibnamefont
  {{Mitra}}}, \bibinfo {author} {\bibfnamefont {Dipongkar}\ \bibnamefont
  {{Talukder}}}, \bibinfo {author} {\bibfnamefont {Sukanta}\ \bibnamefont
  {{Bose}}}, \ and\ \bibinfo {author} {\bibfnamefont {Vuk}\ \bibnamefont
  {{Mandic}}},\ }\bibfield  {title} {\enquote {\bibinfo {title} {{Probing the
  anisotropies of a stochastic gravitational-wave background using a network of
  ground-based laser interferometers}},}\ }\href {\doibase
  10.1103/PhysRevD.80.122002} {\bibfield  {journal} {\bibinfo  {journal}
  {\prd}\ }\textbf {\bibinfo {volume} {80}},\ \bibinfo {eid} {122002} (\bibinfo
  {year} {2009})},\ \Eprint {http://arxiv.org/abs/0910.0858} {arXiv:0910.0858
  [astro-ph.IM]} \BibitemShut {NoStop}%
\bibitem [{\citenamefont {Press}\ \emph {et~al.}(2007)\citenamefont {Press},
  \citenamefont {Teukolsky}, \citenamefont {Vetterling},\ and\ \citenamefont
  {Flannery}}]{Press2007NumericalRT}%
  \BibitemOpen
  \bibfield  {author} {\bibinfo {author} {\bibfnamefont {William~H.}\
  \bibnamefont {Press}}, \bibinfo {author} {\bibfnamefont {Saul~A.}\
  \bibnamefont {Teukolsky}}, \bibinfo {author} {\bibfnamefont {William~T.}\
  \bibnamefont {Vetterling}}, \ and\ \bibinfo {author} {\bibfnamefont
  {Brian~P.}\ \bibnamefont {Flannery}},\ }\bibfield  {title} {\enquote
  {\bibinfo {title} {Numerical recipes: the art of scientific computing, 3rd
  edition},}\ \ }(\bibinfo {year} {2007})\BibitemShut {NoStop}%
\bibitem [{\citenamefont {Bar-Kana}(1994)}]{PhysRevD.50.1157}%
  \BibitemOpen
  \bibfield  {author} {\bibinfo {author} {\bibfnamefont {Rennan}\ \bibnamefont
  {Bar-Kana}},\ }\bibfield  {title} {\enquote {\bibinfo {title} {Limits on
  direct detection of gravitational waves},}\ }\href {\doibase
  10.1103/PhysRevD.50.1157} {\bibfield  {journal} {\bibinfo  {journal} {Phys.
  Rev. D}\ }\textbf {\bibinfo {volume} {50}},\ \bibinfo {pages} {1157--1160}
  (\bibinfo {year} {1994})}\BibitemShut {NoStop}%
\bibitem [{\citenamefont {Regimbau}(2011)}]{Regimbau:2011rp}%
  \BibitemOpen
  \bibfield  {author} {\bibinfo {author} {\bibfnamefont {Tania}\ \bibnamefont
  {Regimbau}},\ }\bibfield  {title} {\enquote {\bibinfo {title} {{The
  astrophysical gravitational wave stochastic background}},}\ }\href {\doibase
  10.1088/1674-4527/11/4/001} {\bibfield  {journal} {\bibinfo  {journal} {Res.
  Astron. Astrophys.}\ }\textbf {\bibinfo {volume} {11}},\ \bibinfo {pages}
  {369--390} (\bibinfo {year} {2011})},\ \Eprint
  {http://arxiv.org/abs/1101.2762} {arXiv:1101.2762 [astro-ph.CO]} \BibitemShut
  {NoStop}%
\bibitem [{\citenamefont {Sandick}\ \emph {et~al.}(2006)\citenamefont
  {Sandick}, \citenamefont {Olive}, \citenamefont {Daigne},\ and\ \citenamefont
  {Vangioni}}]{PhysRevD.73.104024}%
  \BibitemOpen
  \bibfield  {author} {\bibinfo {author} {\bibfnamefont {Pearl}\ \bibnamefont
  {Sandick}}, \bibinfo {author} {\bibfnamefont {Keith~A.}\ \bibnamefont
  {Olive}}, \bibinfo {author} {\bibfnamefont {Fr\'ed\'eric}\ \bibnamefont
  {Daigne}}, \ and\ \bibinfo {author} {\bibfnamefont {Elisabeth}\ \bibnamefont
  {Vangioni}},\ }\bibfield  {title} {\enquote {\bibinfo {title} {Gravitational
  waves from the first stars},}\ }\href {\doibase 10.1103/PhysRevD.73.104024}
  {\bibfield  {journal} {\bibinfo  {journal} {Phys. Rev. D}\ }\textbf {\bibinfo
  {volume} {73}},\ \bibinfo {pages} {104024} (\bibinfo {year}
  {2006})}\BibitemShut {NoStop}%
\bibitem [{\citenamefont {Abbott}\ \emph {et~al.}(2019)\citenamefont {Abbott}
  \emph {et~al.}}]{LIGOScientific:2019vic}%
  \BibitemOpen
  \bibfield  {author} {\bibinfo {author} {\bibfnamefont {B.~P.}\ \bibnamefont
  {Abbott}} \emph {et~al.} (\bibinfo {collaboration} {LIGO Scientific,
  Virgo}),\ }\bibfield  {title} {\enquote {\bibinfo {title} {{Search for the
  isotropic stochastic background using data from Advanced
  LIGO\textquoteright{}s second observing run}},}\ }\href {\doibase
  10.1103/PhysRevD.100.061101} {\bibfield  {journal} {\bibinfo  {journal}
  {Phys. Rev. D}\ }\textbf {\bibinfo {volume} {100}},\ \bibinfo {pages}
  {061101} (\bibinfo {year} {2019})},\ \Eprint
  {http://arxiv.org/abs/1903.02886} {arXiv:1903.02886 [gr-qc]} \BibitemShut
  {NoStop}%
\end{thebibliography}%
\end{document}